\documentclass{amsart}
\usepackage{amssymb,stmaryrd}
\usepackage{amsfonts}
\usepackage{amstext}
\usepackage{algorithmic}
\usepackage{algorithm}
\usepackage{graphicx}
\usepackage{epstopdf}
\usepackage[all]{xy}
\usepackage{MnSymbol}

\numberwithin{equation}{section}

\theoremstyle{definition}

\theoremstyle{remark}

\newcommand{\C}{{\mathbb{C}}}
\newcommand{\Z}{{\mathbb{Z}}}

\newcommand{\<}{{\langle}}

\renewcommand{\>}{{\rangle}}

\newcommand{\wedgeq}{{\wedge\kern-5pt\cdot}}

\newcommand{\tens}{\otimes}

\newcommand{\id}{{\rm id}}

\newcommand{\extd}{{\rm d}}
\newcommand{\del}{{\partial}}
\newcommand{\eps}{\epsilon}

\newcommand{\Vol}{{\rm Vol}}

\begin{document}

\title[Quantum Riemannian geometry and particle creation on the integers]{Quantum Riemannian geometry and particle creation on the integer line}
\keywords{noncommutative geometry,  quantum gravity, lattice gravity,   graph Laplacian, time-dependent harmonic oscillator}

\subjclass[2000]{Primary 81R50, 58B32, 83C57}

\author{Shahn Majid}
\address{Queen Mary University of London\\
School of Mathematical Sciences, Mile End Rd, London E1 4NS, UK}

\email{s.majid@qmul.ac.uk}


\begin{abstract} We construct noncommutative or  `quantum' Riemannian geometry on the integers $\Z$ as a lattice line $\cdots\bullet_{i-1}-\bullet_i-\bullet_{i+1}\cdots$ with its natural 2-dimensional differential structure and metric given by arbitrary non-zero edge square-lengths $\bullet_i{\buildrel a_i\over -}\bullet_{i+1}$. We find for general metrics a unique $*$-preserving quantum Levi-Civita connection, which is flat if and only if $a_i$ are a geometric progression where the ratios $\rho_i=a_{i+1}/a_i$ are constant. More generally, we compute the Ricci tensor for the natural antisymmetric lift of the volume 2-form and find that the quantum Einstein-Hilbert action up to a total divergence is $-{1\over 2}\sum \rho\Delta \rho$ where $(\Delta\rho)_i=\rho_{i+1}+\rho_{i-1}-2\rho_i$ is the standard discrete Laplacian. We take a first look at some issues for quantum gravity on the lattice line. We also examine $1+0$ dimensional scalar quantum theory with mass $m$ and the lattice line as discrete time. As an application, we compute discrete time cosmological particle creation for a step function jump in the metric by a factor $\rho$, finding that an initial vacuum state has at later times an occupancy $\< N\>=(1-\sqrt{\rho})^2/(4\sqrt{\rho})$ in the continuum limit, independently of the frequency.  The continuum limit of the model is the time-dependent harmonic oscillator, now viewed geometrically. \end{abstract}
\maketitle 

\section{Introduction}

The idea that our concept of spacetime should be modified as we approach the Planck scale is now widely accepted, though how precisely to do it is not clear. One popular idea is that spacetime coordinates should  become noncommutative as an expression of quantum gravity corrections, see\cite{Sny,Ma:pla,MaRue,DFR} and our recent work \cite{Ma:sq} for some of the background to this `quantum spacetime hypothesis'. By now there are also several approaches as to how this might be done, such as the  `Dirac operator' (spectral triple) approach of Connes\cite{Con} and, which is the one we use, a constructive approach starting with an abstractly defined `bimodule'  $\Omega^1$ of 1-forms  on the possibly noncommutative coordinate algebra, e.g. \cite{DVM,Mou,BegMa1,BegMa3,Ma:gra,BegMa4}. We refer to \cite{Ma:ltcc}  for an introduction. Some physically interesting models in this bimodule approach were in \cite{BegMa2,MaTao}.  

Another widespread idea for Planck scale spacetime is that there should be some form of discretisation, which again can be done in different ways. The most basic would be to replace spacetime by a lattice or perhaps by a graph and use the methods of discrete geometry in line with lattice field theory and lattice gauge theory. There are also more sophisticated methods such as dynamical triangulations\cite{Loll} and causal set models\cite{Dow}. At least the basic graph approach can be seen as quantum Riemannian geometry\cite{Ma:gra,Ma:sq}, as we explain briefly in the preliminary Section~\ref{secpre}. Here the algebra of functions on a discrete set is commutative but the differential forms $\Omega^1$ are defined by the graph edges and do not commute with functions. 

Although the formalism has been around for a few years now, a downside of the bimodule approach is that the construction of a quantum Levi-Civita connection (QLC) for a chosen quantum metric involves a nonlinear (quadratic) condition which can be solved in individual cases but is hard to solve uniformly across a significant moduli of metrics. This was achieved recently for the square graph, cf \cite{Ma:sq}, and in the present paper our main result at a geometric level is in Section~\ref{secline} to achieve the same for the integers regarded as a line graph. Quantum metrics themselves  are easy to describe for a graph calculus, namely the data is just a `square length' associated to each edge much as in most ideas for lattice approximations. In our case it means a real non-zero number $a_i$ associated to each edge $\bullet_i-\bullet_{i+1}$ of the integer line. Where we part company, however, with conventional discrete or lattice geometry is the concept  of a QLC and hence of curvature, geometric Laplacians (which need not be the usual graph Laplacian) etc., needed for a full picture of quantum Riemannian geometry. This includes an approach to (but not a completely canonical construction of) the Ricci scalar and hence to the Einstein-Hilbert action.

We find, for a general metric on a line graph, a unique QLC that is $*$-preserving. Moreover, this  QLC has curvature as soon as the ratios $\rho_i=a_{i+1}/a_i$ of square-edge-lengths are not constant (a kind of double-derivative of the metric). The reader may wonder how it is possible here that a discrete line could have curvature. The reason is that in fact the intrinsic differential structure of a line is not 1-dimensional. Indeed, on any graph  the number of edges from a node determines the independent directions of travel and for a regular graph the dimension of the cotangent bundle\cite{Ma:gra}. For a line graph there are two directions, up or down in the integer label. There is an associated basis of left-invariant vector fields with respect to the addition structure of $\Z$, namely
\[ \del_\pm=R_\pm-\id\]
where $R_\pm$ are the left and right shift operators. These are related by $\del_-=-R_-\del_+$ so {\em in the continuum limit}, where we can ignore a single shift, we end up with $\del_-=-\del_+$ and only one independent direction for the tangent to a line. However, this is an artefact of the limiting process and more correctly the $\del_\pm$ are related but nevertheless  independent over the coordinate algebra (the corresponding left-invariant 1-forms $e_\pm$ form a global basis of $\Omega^1$). We take the 1-forms $e_\pm$ to anticommute, which implies that $\Omega^2$ is the top degree and is 1-dimensional over the algebra. Thus the intrinsic differential structure of $\Z$ as a line graph is most naturally like that of a `2-manifold'. In effect, the line is thickenned by the discretisation and this is expressed in the quantum geometry. 

After solving the quantum geometry as above, we are then in position to explore a little physics. Here the lattice will be time, so we take a $1+0$-dimensional spacetime point of view. Thus, space is just one point which we do not need to refer to, and functions on spacetime just depend on our discrete time $i\in\Z$. Because the quantum geometry of $\Z$ is naturally 2D, we can have interesting gravity and even quantum gravity. Section~\ref{secmet} takes a first look at this, namely the Einstein-Hilbert action and some issues for its functional integral quantisation (which we do not attempt here). The action turns out to be given by the discrete Laplacian on $\rho_i$ regarded as a positive-valued scalar field. Section~\ref{secscalar} instead looks in detail at the easier case of functional integral quantisation of scalar field theory on $\Z$ of mass $m$. The flat case where $a_i=a$, a constant, is surely not fundamentally new but we give our own take on it including 2-point correlation functions 
\[ \<\phi(i)\phi(j)\>=\imath {D_i D_{n-1-j}\over D_n}\]
for $i<j$ and for the theory restricted to an interval $0,\cdots,n-1$ in $\Z$, where $D_k=\det(B_k- am^2)$ and $B_k$ is the Cartan matrix of $SU_{k+1}$. Here $1/\sqrt{D_k}$ is the partition function for the scalar field theory restricted to an interval of size $k$. In fact, the $D_k$ extend to all real $k$ even though they lose their determinant interpretation when $k$ is not a positive integer, but the formulae then do depend on whether $am^2<4$ or $am^2>4$, of which only the former has the expected continuum limit $a\to 0$. It is not clear if there is physics in the other `deep discrete' phase, but if there is then there would appear to be a phase transition at $am^2=4$.  The continuum phase by contrast is largely similar to the correlators one expects for a Hamiltonian quantisation of the continuum theory $1+0$ theory\cite{Boo} but with an effective mass $m_0$ different from $m$ due to the discretisation.  This is such that $\sin(m_0\sqrt{a}/2)=m\sqrt{a}/2$ in line with indications in other situations such as\cite{Ash,AshBar,FreMa} if we take $\sqrt{a}$ of order the Planck length. 

We then turn to scalar field theory on a general curved metric background, defined by the geometric Laplacian associated to the QLC. Using this, we consider  wave propagation through a time interval $0,\cdots,n-1$ when the metric fluctuates, with the metric constant and before and after. We illustrate the method on the case of a step function, where the metric jumps from a constant $a$ to a constant $b$, and a `bump' where it then immediately jumps back to $a$. In both cases, we do a preliminary calculation of a kind of `cosmological particle creation'\cite{ParNav} due to this metric change in which the vacuum seen by observers before the metric change has an occupation number $\<N\>$ as seen by later observers. This is a similar underlying effect as in Bekenstein-Hawking radiation but applies more generally\cite{MukWin} and has an earlier origin,  notably in the work of Parker. In particular, what we find is not thermal in that the effect in the continuum limit $a\to 0$ will be independent of the mass-frequency and of the shape of the step. These calculations are plausible but ad-hoc  in so far as we will use conventional Hamiltonian quantisation to treat the discrete quantum theory rather than a systematic general framework. We also  normalise the in and out waves by requiring unitarity of the Bogoliubov transformation rather than by more geometric arguments based on conserved currents. Our results do, however, provide a proof of concept as well as a new `curved time' point of view of the time-dependent harmonic oscillator. The latter in the continuum has been extensively studied since the early works \cite{Lew,LR} and also plays a role in more recent works\cite{BFV}. The paper ends with some concluding remarks.

\subsubsection*{Acknowledgements} I would like to thank C. Fritz for some very helpful discussions about Hawking radiation and cosmological particle creation.

\section{Preliminaries: quantum geometric formalism}\label{secpre}

An introduction to the formalism of the  constructive `bottom up' approach to quantum Riemannian geometry\cite{BegMa1,BegMa2,BegMa3,BegMa4,Ma:gra,MaTao}  that we will use is in \cite{Ma:ltcc} and  in our recent work\cite{Ma:sq}, so here we will say just enough to be self-contained. We will only need the case of graphs  from\cite{Ma:gra} and moreover only the discrete group case.   It is important, however, that our constructions are not ad-hoc to the extent possible but part of a general framework that equally well specialises to classical Riemannian geometry and many other cases of interest. 

The general theory is based on an algebra $A$ of `functions on spacetime' which could be noncommutative but which in our case will be the commutative algebra $C(X)$ of complex functions on a discrete set $X$ with pointwise product. We next need a `differential structure' in the form of an $A-A$ bimodule $\Omega^1$ of `1-forms'. A bimodule just means that we can associatively multiply 1-forms by functions from the left or the right. We also need an `exterior derivative' $\extd:A\to \Omega^1$ obeying the Leibniz rule with respect to these left and right products. In our case, this data amounts to a directed graph with vertex set $X$ and $\Omega^1$ a vector space with basis $\{\omega_{x\to y}\}$ labelled by arrows of the graph. The bimodule and exterior derivative structures are
\[  f.\omega_{x\to y}=f(x)\omega_{x\to y},\quad \omega_{x\to y}.f=f(y)\omega_{x\to y},\quad \extd f=\sum_{x\to y}(f(y)-f(x))\omega_{x\to y}.\]
We will be interested in the case where the graph is bidirected i.e., for every arrow $x\to y$ there is an arrow $y\to x$. In other words, the data is just a usual undirected graph which we understand as arrows both ways in the above formulae. A metric is a tensor $g\in \Omega^1\tens_A\Omega^1$ which is nondegenerate in a certain sense. In our case of a graph calculus it amounts to 
\[ g=\sum_{x\to y}g_{x\to y}\omega_{x\to y}\tens\omega_{y\to x}\in \Omega^1\tens_{C(X)}\Omega^1\]
for nonzero weights $g_{x\to y}$ for every edge. There are two different notions of symmetry for the metric, one is functorial i.e., works for any algebra with calculus once we have extended $\Omega^1$ to higher forms $\Omega$ generated by $\Omega^1$ and $A$ and with $\extd$ extended by the super-Leibniz rule and such that $\extd^2=0$. In that context a metric is  `quantum symmetric' if $\wedge(g)=0$ for the wedge product of $\Omega$. The other is specific to graphs and we will say that $g$ is {\em edge-symmetric} if 
\begin{equation}\label{edgesym}g_{x\to y}=g_{y\to x}
\end{equation}
 (in principle, the `square length' associated to an edge could depend on the direction of travel; we assume it does not). Although not part of the general theory, we will see that this variant works better when we apply it to the line graph. 

Next we want a Riemannian connection $\nabla:\Omega^1\to \Omega^1\tens_A\Omega^1$ where classically the left hand copy of the output would contract against a vector field to give a covariant derivative $\Omega^1\to \Omega^1$ (but we do not need these themseleves). $\nabla$ should obey a pair of Leibniz rules, the one for $\nabla(f\omega)$ is the usual one and the one for $\nabla(\omega f)$ requires the existence of a bimodule map $\sigma:\Omega^1\tens_A\Omega^1\to \Omega^1\tens_A\Omega^1$ called the `generalised braiding' c.f.\cite{DVM,Mou}. This is then used to extend the action of $\nabla$ to a connection on $\Omega^1\tens_A\Omega^1$, which then allows us to write down metric compatibility as $\nabla g=0$. For $\nabla$ to be torsion free similarly makes sense abstractly, as the condition $\wedge\nabla=\extd$. When both of these hold, we have a `quantum Levi-Civita connection' or QLC. We omit all the details in favour of the following result for graph calculi (and other `inner' ones) that every connection has the form\cite[Thm.~2.1]{Ma:gra}
\[ \nabla\omega=\theta\tens\omega-\sigma_\theta(\omega)+\alpha(\omega);\quad \sigma_\theta(\omega)=\sigma(\omega \tens\theta)\]
where in the graph case $\theta=\sum_{x\to y}\omega_{x\to y}$ and where $\alpha:\Omega^1\to \Omega^1\tens_A\Omega^1$ and $\sigma$ are bimodule maps (they commute with products by functions from either side). 
In this case, vanishing torsion and metric compatibility respectively become\cite{Ma:gra}
\begin{equation}\label{torsionfree}  \wedge (\id+\sigma)=0,\quad \wedge\alpha=0\end{equation}
\begin{equation}\label{mc} \theta\tens g+(\alpha\tens\id)g-\sigma_{12}(\id\tens(\alpha-\sigma_\theta)g=0\end{equation}
In the graph case we are often forced to have $\alpha=0$, in which case we are just solving for $\sigma$. 

For every connection, we have a Riemannian curvature $R_\nabla=(\extd\tens\id-\id\wedge\nabla)\nabla:\Omega^1\to \Omega^2\tens_A\Omega^1$. Ricci requires more data and the current state of the art (but probably not the only way) is to introduce a lifting map $i:\Omega^2\to\Omega^1\tens_A\Omega^1$. Applying this to the left output of $R_\nabla$, we are then free to `contract' by using the metric and inverse metric $(\ ,\ )$  to define ${\rm Ricci}\in \Omega^1\tens_A\Omega^1$ \cite{BegMa2}. The Ricci scalar is then $S=(\ ,\ ){\rm Ricci}\in A$.  More canonically, we have a geometric quantum Laplacian $\Delta=(\ ,\ )\nabla\extd: A\to A$ defined again along lines that generalise the classical concept to any algebra with differential structure, metric and connection. 

For physics, we work over $\C$ with $A$ a $*$-algebra, which in our case of $C(X)$ is by pointwise complex conjugation. We want this to be compatible with $\Omega^1$, which in our case amounts to $\omega_{x\to y}^*=-\omega_{y\to x}$ and then implies that $\extd$ commutes with $*$. We also want the metric to be compatible with $*$ in the sense $g^\dagger=g$ (where $\dagger$ means to apply $*$ to each factor and flip the two factors), which in our case amounts to $g_{x\to y}$ real. An edge will be called {\em timelike} if this is positive, its interpretation being the square of the edge length under the metric. Finally, we want the connection to be $*$-preserving which in the case of interest comes down to  
\begin{equation}\label{*conn}(\dagger\circ\sigma)^2=\id,\quad \sigma\circ\dagger\circ\alpha=\alpha *.\end{equation}

\section{QLCs for metrics on $\Z$}\label{secline}

The  conditions for a QLC depend on how $\Omega^2$ is defined, but for graph calculi there is a canonical choice of this in the case when $X$ is a group and the graph a Cayley graph generated by right translation by a set of generators. Here the edges are of the form $x\to xa$ where $a$ is from the generating set and the product is the group product. In this case there is a natural basis of left-invariant 1-forms $e_a=\sum_{x\to xa} \omega_{x\to xa}$. These obey the simple rules 
\[ e_a f= R_a(f)e_a,\quad \extd f=\sum_a (\del_a f)e_a,\quad \del_a=R_a-\id,\quad R_a(f)(x)=f(xa)\]
defined by the right translation operators $R_a$ as stated. In this case  $\Omega$ is canonically generated by the $e_a$ with certain `braided-anticommutation relations' c.f. \cite{Wor}. In the case of an Abelian group (which is all we will need) this is just the usual  Grassmann algebra on the $e_a$, i.e., they anticommute. 

Now let $X=\Z$ with generators $\{+1,-1\}$. Then the Cayley graph is the integer line $\cdots\bullet_{i-1}-\bullet_i-\bullet_{i+1}\cdots$. There are two left-invariant forms 
\[ e_+=\sum_i \omega_{i\to i+1},\quad e_-=\sum_i\omega_{i\to i-1}\]
with $e_\pm f=  R_\pm(f) e_\pm$, $\extd f=\sum_\pm(\del_\pm f)e_\pm$ where $\del_\pm =R_\pm-\id$ and $R_\pm(f)(i)=f(i\pm 1)$. The exterior algebra has $e_\pm$ anticommuting and $e_+^*=-e_-$. For the QLC we use \cite[Thm.~2.1]{Ma:gra} and are forced to have the map $\alpha(e_\pm)=0$. We fix this from now and will use $\alpha$ for other purposes. Similarly, the most general form of braiding bimodule map $\sigma$ obeying the torsion free condition (\ref{torsionfree}) is forced to be of the form
\[ \sigma(e_+\tens e_+)=\alpha e_+\tens e_+,\quad \sigma(e_-\tens e_-)=\beta e_-\tens e_-\]
\[ \sigma(e_+\tens e_-)=(\gamma+1) e_-\tens e_++ \gamma e_+\tens e_-,\quad \sigma(e_-\tens e_+)=\delta e_-\tens e_++ (\delta+1) e_+\tens e_-\]
where $\alpha,\beta,\gamma,\delta$ are functional coefficients. The `inner' element is $\theta=e_++e_-$ so the corresponding connection $\nabla e_\pm= (e_++ e_-)\tens e_\pm - \sigma_\theta(e_\pm)$ has 
\[ \sigma_\theta(e_+)=\sigma(e_+\tens (e_+\tens e_-))=\alpha e_+\tens e_++\gamma e_+\tens e_-+(\gamma+1)e_-\tens e_+\]
\[ \sigma_\theta(e_-)=\sigma(e_-\tens(e_+\tens e_-))=\beta e_-\tens e_-+\delta e_-\tens e_++(\delta+1)e_+\tens e_-,\]
giving
\[ \nabla e_+=(1-\alpha)e_+\tens e_+-\gamma(e_+\tens e_-+e_-\tens e_+),\quad \nabla e_-=(1-\beta)e_-\tens e_--\delta(e_+\tens e_-+e_-\tens e_+).\]
To be $*$-preserving we need (\ref{*conn}), which  comes out as 
\begin{equation}\label{starc} \alpha R_+^2\bar\beta=1,\quad \delta\bar\delta+ \delta+\bar\gamma(\delta+1)=0,\quad \gamma\bar\gamma+ \gamma+\bar\delta(\gamma+1)=0.\end{equation}
To be metric compatible, we need (\ref{mc}) without the map $\alpha$ there, still to be analysed.

\subsection{QLCs in the quantum-symmetric case.} The general form of real quantum-symmetric metric is 
\[ g=a (e_+\tens e_-+e_-\tens e_+)=\sum_i a_i (\omega_{i\to i+1}\tens\omega_{i+1\to i}+\omega_{i\to i-1}\tens\omega_{i-1\to i})\]
where $a_i:=a(i)$ are real and non-zero. To have a fixed signature, we assume they are all $>0$. Note that  $g_{i\to i+1}=a_i= g_{i\to i-1}$ so the above quantum-symmetric metric is not quite edge-symmetric unless $a$ is constant. The inverse metric is 
\[ (e_\pm,e_\pm)=0,\quad (e_+,e_-)={1\over R_+(a)},\quad (e_-,e_+)={1\over R_-(a)}.\]
Writing  out  (\ref{mc}) and matching the coefficients in the triple tensor product basis, we obtain six equations
\[ R_+a=a(R_+\delta+1)\alpha=a\gamma R_+\delta+a(\delta+1)R_-\alpha,\quad R_-a=a(R_-\gamma+1)\beta=a\delta R_-\gamma+a(\gamma+1)R_+\beta\]
\begin{equation}\label{sol2} (\gamma+1)R_+\delta+\delta R_-\alpha=0,\quad (\delta+1)R_-\gamma+\gamma R_+\beta=0.\end{equation}
Using the second set, and writing $\rho=R_+a/a$, we write two of the first line as
\[ \rho=R_-\alpha- R_+\delta,\quad {1\over R_-\rho}=R_+\beta-R_-\gamma\]
(which are the `cotorsion equations' in the general theory in \cite{Ma:gra}). The other two equations in that line can be written as
\begin{equation}\label{sol1} \delta={R_-\del_-\alpha\over 1+R_-\alpha},\quad \gamma={R_+\del_+\beta\over 1+R_+\beta}\end{equation}
so that there are only two free functions $\alpha,\beta$. Putting the form of $\delta,\gamma$ into the cotorsion equations gives 
\begin{equation}\label{sol3} R_-\alpha=\rho-1+{\rho\over\alpha},\quad R_+\beta={1\over \beta R_-\rho}+ {1\over R_-\rho}-1\end{equation}
as recursion equations which we use to solve the system given $\rho$ and initial conditions. We then define $\delta,\gamma$ by (\ref{sol1}) and need to verify if (\ref{sol2}) then hold.  Thus, a QLC may not always exist if the last step fails for all choices of initial conditions on $\alpha,\beta$.

We apply the above to the example of $\rho=c$ a non-zero constant, so $a(i)=c^i a(0)$ grows or decays exponentially. Then there are two solutions for QLCs:
\[ (i)\quad \alpha=c,\quad \beta(i)=\begin{cases}q &  i\ {\rm even}\\ {q^{-1}+ 1-c\over  c} & i\ {\rm odd}\end{cases},\quad \gamma(i)=\begin{cases}{q^{-1}- c\over c-1-q^{-1}} &  i\ {\rm even}\\ {c(1-qc)\over (1-c)(1-qc)+q} & i\ {\rm odd}\end{cases},\quad\delta=0\]
\[ (ii)\quad \alpha(i)=\begin{cases} q &  i\ {\rm even}\\ {  c \over q+1-c  } & i\ {\rm odd}\end{cases},\quad \beta={1\over c},\quad \gamma=0,\quad \delta(i)=\begin{cases} {q-c\over q(c-1)+c} &  i\ {\rm even}\\  q^{-1}c-1 & i\ {\rm odd}.\end{cases}\]
given as solutions for the generalised braiding $\sigma$. Of these, there is a unique $*$-preserving QLC namely $q=c^{-1}$ in case (i) and $q=c$ in case (ii), giving in both cases $\alpha=c, \beta=1/c$ and $\gamma=\delta=0$. This has
\begin{equation}\label{QLCcon} \nabla e_+=(1-c) e_+\tens e_+,\quad \nabla e_-=(1-{1\over c})e_-\tens e_-\end{equation}
with zero curvature. So the constant $\rho$ case leads to a unique connection and it is flat. 

For a generic quantum symmetric metric, however, (\ref{sol2}) will not hold and we will not have a QLC. In this case, we can work with the much larger moduli of cotorsion-free  $*$-preserving connections or with metric compatible $*$-preserving connections with torsion. Or, there is a nice alternative in our case of a graph calculus, which we look at next.

\subsection{QLCs in the edge-symmetric case}\label{secQLCsym}

Given the paucity of solutions in the preceding section, we now look at the case where $g$ is not assumed quantum symmetric but instead edge-symmetric, and repeat the above steps. In this case we need 
\[ g=a e_+\tens e_-+ R_-a e_-\tens e_+=a e_+\tens e_-+e_- a\tens e_+\]
arranged so that $a(i)=g_{i\to i+1}=(R_-a)(i+1)=g_{i+1\to i}$ at all $i$ as the most general form of edge-symmetric metric, for any  real non-zero function $a$. The inverse metric is
\[ (e_\pm,e_\pm)=0,\quad (e_+,e_-)={1\over a},\quad (e_-,e_+)={1\over R_-a}.\]

In this case similar computations for the basis coefficients in the tensor power give metric compatibility (\ref{mc}) as
\[ (R_+\delta+1)\alpha=\rho,\quad \beta(R_-\gamma+1)={1\over R^2_-\rho}\]
\[ 1= (R_+\delta)\gamma+ R_-({\alpha\over\rho})(\delta+1),\quad 1=(R_-\gamma)\delta+(R_-\rho)(R_+\beta)(\gamma+1)\]
\[ (R_+\delta)(\gamma+1)+R_-({\alpha\over\rho})\delta=0,\quad (R_-\rho) (R_+\beta)\gamma+(R_-\gamma)(\delta+1)=0.\]
Taken together, these are equivalent to 
\begin{equation}\label{newsol1} \delta=R_-({\rho\over\alpha})-1,\quad \gamma={1\over (R_-\rho)R_+\beta}-1,\quad ((R_-\rho)R_+\beta-1)(\alpha-\rho)=0\end{equation}
\begin{equation}\label{newsol2} \alpha-\rho=-\alpha(R_+\beta)R_-(\alpha-\rho),\quad  R_+\beta-{1\over R_-\rho}=-\alpha (R_+\beta)R_+(R_+\beta- {1\over R_-\rho}).\end{equation}
Now if $\alpha(i)=\rho(i)$ at some $i$ then by iterating the first of (\ref{newsol2}), we will be forced into the following case
\[{\rm Case\ (i):}\quad \alpha=\rho,\quad \beta R_+\beta={1\over (R_-^2\rho)R_-\rho},\quad\gamma={1\over (R_-\rho)R_+\beta}-1,\quad \delta=0\]
with general solution
\[ \alpha=\rho,\quad \beta(i)={(q \rho(-2))^{(-1)^i}\over \rho(i-2)},\quad\gamma=(q \rho(-2))^{(-1)^i}-1,\quad \delta=0\]
where $q$ is a free parameter. Similarly, if $\beta(i+1)=1/\rho(i-1)$ at some $i$ then we will be forced into 
\[ {\rm Case\ (ii):}\quad \alpha R_+\alpha=\rho R_+\rho,\quad \beta={1\over R_-^2\rho},\quad\gamma=0,\quad \delta=R_-({\rho\over\alpha})-1\]
with general solution  
\[ \alpha(i)=\left({q\over\rho(0)}\right)^{(-1)^i}\rho(i),\quad \beta={1\over R_-^2\rho},\quad\gamma=0,\quad\delta=\left({q\over\rho(0)}\right)^{(-1)^i}-1\]
where again $q$ is a free parameter. These are the only possibilities since
by the 2nd of (\ref{newsol1})  at $i=0$ (say) we must be in one case or the other. 

Among the above solutions (i) and (ii), the $*$-preserving condition (\ref{starc}) forces them both to be the same solution
\[ \alpha=\rho,\quad\beta={1\over R_-^2\rho},\quad\gamma=\delta=0\]
(setting $q=1/\rho(-2)$ in the first solution or $q=\rho(0)$ in the second). For constant $\rho$ we get the same connection (\ref{QLCcon}) as before but now we have a unique $*$-preserving QLC  for any general edge-symmetric metric. The connection and curvature for this are
\begin{equation}\label{edgenable} \nabla e_+=(1-\rho)e_+\tens e_+,\quad \nabla e_-=(1-{1\over R_-^2\rho})e_-\tens e_-\end{equation}
\begin{equation}\label{edgecurv} R_\nabla e_+=\del^-\rho \Vol\tens e_+,\quad R_\nabla e_-=-\del^+\left({1\over R_-^2\rho}\right)\Vol\tens e_-\end{equation}
where $\Vol=e_+\wedge e_-$ is the top form and
\[ \del^-\rho={a\over R_-a}-{R_+a\over a}={\tau\over a R_-a},\quad \del^+\left({1\over R_-^2\rho}\right)={R_-\tau\over a R_+a}   ;\quad \tau:=a^2-(R_+a)R_-a.\]
This edge-symmetric metric case seems much more reasonable in light of the unique $*$-preserving QLC. We proceed exclusively in this case. 

\subsection{Ricci scalar and Einstein-Hilbert action for edge-symmetric metrics}\label{secmet}

For the unique connection at the end of Section~\ref{secQLCsym},  the Ricci tensor is defined by the contraction
\[ {\rm Ricci}=((\ ,\ )\tens\id)(\id\tens i\tens\id)(\id\tens R_\nabla)g\]
for suitable lift map $i:\Omega^2\to \Omega^1\tens_A\Omega^1$ which for our Grassmann exterior algebra case has the canonical choice $i(\Vol)={1\over 2}(e_+\tens e_--e_-\tens e_+)$. This gives
\[ {\rm Ricci}={1\over 2}\left( \del^+\left({1\over R_-\rho}\right)e_+\tens e_-+ \del^- R_-\rho e_-\tens e_+\right)={1\over 2}\left(  {\tau\over (R_+a) R_+^2a} e_+\tens e_-+  {R_-\tau\over (R_-a)R_-^2a} e_-\tens e_+\right)\]
and Ricci scalar
\begin{equation}\label{RicciS} S={1\over 2a}\left( -\del^-\left({1\over\rho}\right)+R_-(\rho\del^-\rho)  \right)={1\over 2}\left({\tau\over a (R_+a) R_+^2a}+R_- \left({\tau\over a^2 R_- a}\right)\right).\end{equation}
Note that the contraction conventions and the half from $i$ mean that our Ricci is - 1/2 of the usual Ricci in the classical limit and more standard conventions.

As a measure, we take $\mu=a$ (in the commutative case this would be $\sqrt{-\det(g)}$ for our form of metric) and discard the total divergence up to a constant (assuming $\rho$ rapidly approaches constant values at large $i$). Then 
\begin{align*}S_g&=-2\sum_\Z\mu S={\rm Const.}-\sum_\Z \rho\del^-\rho={\rm Const.}-\sum_i \rho(i)(\rho(i-1)-\rho(i))\\
&={\rm Const.}-{1\over 2}\sum_i \rho(i)(\rho(i+1)+\rho(i-1)-2\rho(i))={\rm Const.}-{1\over 2}\sum_\Z\rho\Delta_\Z \rho\end{align*}
where the $-2$ compensates for our Ricci convention and we added a further total divergence $\del^+(\rho R_-\rho)$ to get the symmetric form. 
The first form resembles  some kind of `Dirac operator' on the relative metric variation function $\rho$. The second form is the action for the usual discrete Laplacian 
\begin{equation}\label{DZ} (\Delta_\Z\rho)(i)=\rho(i+1)+\rho(i-1)-2\rho(i)\end{equation} for a scalar field on $\Z$. 

Our interpretation, however, is different as the field is the metric $a$ not the relative ratios  $\rho$ 
 (which play a role as a kind of `derivative'). For the functional integral, in order not to be completely formal, we limit ourselves to configurations which vary only in a finite subset, say $a_0,\cdots,a_n$ so that $a(i)=a_i$ and $\rho(i)=\rho_i$ are sequences of the form 
 \[ a=(\dots,a_0,a_0,a_0,a_1,a_2,\cdots,a_n,a_n,a_n,\cdots),\quad  \rho=(\dots,1,1,{a_1\over a_0},{a_2\over a_1},\cdots,{a_n\over a_{n-1}},1,1,\dots)\]
 or conversely, given $\rho_0,\cdots,\rho_{n-1}$ and fixed  $a_0=q$ say (to be concrete, it plays a role like a constant of integration) the corresponding $a_1,\cdots,a_n$ are
 \[ a_1=q\rho_0,\quad a_2=q\rho_0\rho_1,\quad ...\quad a_n=q \rho_0\cdots\rho_{n-1}.\]
 Note that we are fixing just the values for large negative $i$ up to and including $a_0$ and allowing the values for large positive $i$ to float as the last value $a_n$. If one fixed the latter to also to be $q$ (say) then our equivalent description would need one more variable $\rho_n$ with the constraint $\rho_0\cdots\rho_n=1$ which is harder to analyse and does not seem more physical. Dropping any constants  in $S_g$, we have
 \begin{align*} S_g&=-{1\over 2}\big(\rho_0(\rho_1+2-2\rho_0)+\rho_1(\rho_2+\rho_0-2\rho_1)+\cdots +\rho_{n-2}(\rho_{n-1}+\rho_{n-3}-2\rho_{n-2})\\
 &\qquad+\rho_{n-1}(2+\rho_{n-2}-2\rho_{n-1})\big)\\
 &=\sum_{i=0}^{n-1}\rho_i^2-(\rho_0+\rho_0\rho_1+\rho_1\rho_2+\cdots +\rho_{n-2}\rho_{n-1}+\rho_{n-1})\\
 &=\sum_{i=0}^{n-1}r_i^2-(r_0r_1+r_1r_2+\cdots+r_{n-2}r_{n-1})={1\over 2} r^T B_n r\end{align*}
where in the last line we shifted to the relative metric differentials
\[ r_i=\rho_i-1=a_i^{-1}\del^+ a_i\]
and dropped a constant term. We work with this quadratic action in practice, with the underlying bilinear form $B_n$ given by the Cartan matrix of the Lie algebra $su_{n+1}$.  This and its eigenvalues $\lambda_j$ from \cite{An}   are
\begin{equation}\label{Bn}B_n=\begin{pmatrix}2 &-1&0 & 0 &\cdots & 0\\ -1 & 2 & -1&0&\cdots& 0\\ &\vdots & & &  \cdots & \\ 
0 & \cdots  & 0& -1& 2&-1\\
0&\cdots &0 & 0&-1&2\end{pmatrix},\quad \lambda_j=4\sin^2\left({\pi j\over 2(n+1)}\right),\quad j=1,\cdots,n.\end{equation}

For the functional integral we still need to choose a measure on the field space and this depends on what we consider our primary variables. If as usual we take the metric coefficients $a$ then we need the Jacobean for the change $a_1,\cdots,a_n$ to $\rho_0,\cdots,\rho_{n-1}$,
\[ J=|{\del a\over\del\rho}|=q^{n+1} \rho_0^{n}\rho_1^{n-1}\cdots \rho_{n-1}= a_0a_1\cdots a_n.\]
with the range $(0,\infty)$ for the $\rho_i$ if all the $a_i$ are in this range. The partition function with measure $\extd a_1\cdots\extd a_n$ becomes in the new variables
\[ Z=\int_0^\infty \extd \rho_0\cdots \extd\rho_{n-1}\, \rho_0^n\rho_1^{n-1}\cdots\rho_{n-1}e^{{\imath\over G}S_g}\]
where we have inserted a dimensionless coupling constant $G$, the $\rho_i$  also being dimensionless. If, on the other hand, we consider the $\rho$ as the primary fields then we can omit the $\rho$ powers in the measure.  

For example, $Z$ without the $\rho_i$ factors converges to Fresnel C and S functions of $G$ in the simplest case $n=1$. We do this with the $r_i$ variables (which means dropping a constant phase in $Z$) then for $n=1$,
\[Z= \int_{-1}^\infty\extd r_0e^{{\imath \over G}r_0^2}=\sqrt{G \pi\over 2} \left(\frac{1}{2}+C\left(\sqrt{2\over G\pi }\right)+\imath \left(\frac{1}{2}+S\left(\sqrt{2\over G\pi }\right)\right)\right)\sim e^{{\pi\imath\over 4}}\sqrt{G \pi\over 4}\]
for large $G$ and twice the same asymptotic form in the other limit $G\to 0$. 
For $n=2$, we do not have a closed form but one can show similarly that
\[ Z=\int_{-1}^\infty\extd r_0\extd r_1e^{{\imath \over G}( r_0^2+r_1^2-r_0 r_1)}\sim {1\over 2}\sqrt{G \pi\over 2}+\imath\frac{2 \pi  G}{3 \sqrt{3}}\]
for large $G$ and again tending to zero for small $G$. In both cases, if we only integrate $r_i$ from zero then the results would be simpler with the large $G$ asymptotic form now applying exactly for all $G$. The integrals with powers of $\rho_i$ generally do not converge and would need more care to make sense of. We will look at this in detail in the simpler case of the scalar field in the next section, and come back to the quantum gravity theory elsewhere; one would need more insight as to  the correct measure and range of integration. 

\section{Scalar fields on constant and curved edge-symmetric backgrounds}\label{secscalar}

Also associated to the quantum Riemannian geometry in Section~\ref{secQLCsym} is a geometric Laplacian $\Delta =(\ ,\ )\nabla\extd $ which comes out as 
\[ \Delta \phi=(\ ,\ )\nabla(\del^+\phi e_++\del^-\phi e_-)=\left({1\over a}+{1\over R_-a}\right)\del^-\del^+\phi={1+R_-\rho\over a}\del^-\del^+\phi=-{1+R_-\rho\over a}\Delta_\Z\phi \]
where $\Delta_\Z$ is the usual discrete Laplacian or double-differential on $\Z$ as in (\ref{DZ}). With the same measure $\mu=a$ as before, we have action
\[ S_\phi={1\over 2}\sum\mu \bar\phi({\Delta\over 2}-m^2)\phi=\sum_i{1+\rho_{i-1}\over 4}\bar\phi_i(2\phi_i-\phi_{i+1}-\phi_{i-1})-{1\over 2}\sum_i a_im^2|\phi_i|^2\]
where $\phi_i=\phi(i)$. Here the metric coefficient $a_i=a(i)$ has inverse mass-square dimension and effectively scales the mass term, while the metric ratio `difference' $\rho$ comes into the kinetic term. We take $\Delta/2$ in the action (not $-\Delta$) in order to have a time-like (positive sign) double differential for the wave equation in the constant metric continuum limit and the standard normalisation. 

\subsection{Scalar fields with constant metrics} \label{secconstant}

For a constant metric $\rho=1$, we reduce to the standard discrete wave equation and action
\[ ({1\over a}\Delta_\Z+  m^2)\phi=0,\quad S_\phi=-{1\over 2}\sum\bar\phi(\Delta_\Z+a m^2)\phi\]
 on a 1D lattice. We study this case for reference (it can be treated in several well-known ways). In fact, the analysis for the quantum theory is identical to that for $\rho$ in the preceding section but now $\phi$ is not limited to positive real values. For example, if we consider compact support 
\[\phi=(\cdots,0,0,\phi_0,\cdots,\phi_{n-1},0,0,\cdots)\]
then the action reduces  to
\[ S_\phi={1\over 2}\left(\bar\phi B_n\phi - am^2\bar\phi\phi\right)\]
where $B_n$ is the $su_{n+1}$ Cartan matrix. Then
\[ Z=\int_{-\infty}^\infty\extd\phi_0\cdots\extd\phi_{n-1}\extd\bar\phi_0\cdots\extd\bar\phi_{n-1}e^{{\imath\over \beta} S_\phi}=\pm{\left({2\pi\imath\beta}\right)^n\over D_n}.\]
Here $\beta$ is a dimensionless coupling constant since our model is more like a 2D model with the scalar field dimensionless, and the functional integral only makes sense after perturbation of $B_n$ by $\imath\eps$, $\eps>0$, to give the result shown with 
\begin{equation}\label{Dn} D_n=\det(B_n-a m^2)=\sum _{k=0}^n \left(-a m^2\right)^k \binom{k+n+1}{2 k+1}.\end{equation}
The correlation functions, defined in the same way but with products of fields in the integrand and divided by $Z$ are, for $n=2$, 
 \[ \<\phi_i\>=\<\phi_i\phi_j\>=0,\quad \<\bar\phi_i\phi_i\>=-{2\imath (2-a m^2)\over (1-am^2)(3-a m^2)},\quad
  \<\bar\phi_0\phi_1\>=-{2\imath\over (1-am^2)(3-a m^2)}.\]
 For real scalar fields there is a similar story since $B_n$ can be taken with real eigenvectors. Then after diagonalising, each integration gives us the square root of an eigenvalue and hence the square root of the previous $Z$ up to a possibly fractional power of $i$. For example, if $a m^2<<1$ then the theory with real $\phi_0,\cdots,\phi_n$ has partition function
 \[ Z=\int_{-\infty}^\infty\extd\phi_0\cdots\extd\phi_{n-1}e^{{\imath\over \beta} S_\phi}={(2\pi\imath \beta)^{n\over 2}\over \sqrt{D_n}}.\]
 This assumes that $a m^2$ is not an actual eigenvalue of $B_n$ otherwise $D_n$ as its characteristic polynomial vanishes.  For $n=2$ the correlators are then $-1/2$ of the previous $\bar\phi\phi$  case. For $n=3$, 
   \[ \<\phi_0^2\>=\<\phi_2^2\>=\imath{(1-am^2)(3-am^2)\over(2-am^2)(2- 4 a m^2+a^2 m^4)},\quad\<\phi_1^2\>=\imath{2-am^2\over 2- 4 a m^2+a^2 m^4},\]
  \[ \<\phi_0\phi_1\>=\<\phi_1\phi_2\>={\imath\over 2- 4 a m^2+a^2 m^4},\quad \<\phi_0\phi_2\>={\imath\over(2-am^2)(2- 4 a m^2+a^2 m^4)}.\]
The general pattern in the $\phi_0,\cdots,\phi_{n-1}$ theory for correlation functions appears to be
\[ \<\phi_i\phi_j\>=\<\phi_j\phi_i\>=\imath{D_i D_{n-1-j}\over D_n},\quad  i\le j.\]
This has been verified for small $n$ (rather than writing out a formal proof). We now consider the limit $n\to \infty$ and to do this we note that $D_i$ can be summed  and the resulting expression makes sense for all $i$ including non-integer and negative values. We analyse this in two cases.

(i) Continuum phase: $a m^2<4$. In this case we write
\[ 2- a m^2 + m a^2\sqrt{am^2-4}=2 e^{\imath x};\quad 2\sin(x)=   \sqrt{am^2 \left(4-a m^2\right)},\quad 2\cos(x)=2-a m^2\]
for some phase angle $x>0$. Then one can verify that
\[ D_i={\sin(x (i+1))\over \sin(x)}\]
which has a limit $i+1$ as $a m^2\to 0$ and hence $x\to 0$ (the continuum limit). One can check that $D_i$ obeys the original scalar field wave equation on $\Z$, 
\[ (\Delta_\Z D)_i= D_{i+1}+D_{i-1}-2 D_i=2(\cos(x)-1)D_i=- a m^2 D_i\]
and from this one can show that the Green function, defined as $\imath$ times the 2-point correlation function, inverts the wave operator (summing over $k$), 
\begin{equation}\label{green} (\Delta_\Z+a m^2)_{i k}\<\phi_k\phi_j\>= \<((\Delta_\Z+a m^2)\phi)_i\phi_j\>=-\imath\delta_{ij}.\end{equation}
Here the $i\ne j$ case follows from the wave equation for one or other of the numerator $D$ factors, but at $i=j$ we have, due to the normal ordering and the wave equation,
\begin{align*}
 \<(\phi_{i+1}&+\phi_{i-1}+(am^2-2)\phi_i)\phi_i\>={\imath\over D_n}(D_iD_{n-2-i}+(D_{i-1}+(am^2-2) D_i)D_{n-1-i})\\
 &={\imath\over D_n}(D_iD_{n-2-i}-D_{i+1}D_{n-1-i})=-\imath.\end{align*}

We also note that the correlation function in the limit where $x\to 0$ between a field near the boundary (such as $\phi_0$ at the left one) and fields  `in the bulk' (meaning at a fixed fraction of $n$ as $n\to \infty$) are finite in this limit. For example
\[ \<\phi_0\phi_i\>=\imath {D_{n-1-i}\over D_{n}}\to \imath,\quad \<\phi_0\phi_{n-1\over d}\>=\imath {D_{(n-1)(1-{1\over d})}\over D_n}\to  \imath\left(1- {1\over d}\right)\]
if we first set $x\to 0$ and then set $n\to\infty$ (with $n-1$ a multiple of $d$). This decreases as we move through more and more of the bulk eventually to  $\<\phi_0\phi_{n-1}\>\to 0$ for the correlation from near one boundary to the other. By contrast, the correlator diverges between two fields in the bulk, for example
\[ \<\phi_{n-1\over 2}^2\>=\imath {D_{n-1\over 2}^2\over D_n}=\imath{\tan(x ({n+1\over 2}))\over \sin(x)}\to\imath\infty\]
in the same double limit. 

(ii) Discrete phase  $a m^2>4$.  In this case we let $y>0$ be defined by either of 
\[  2-a m^2\mp \sqrt{a m^2 \left(a m^2-4\right)}= -2 e^{\pm y};\quad 2 \sinh(y)=\sqrt{a m^2(a m^2-4)},\quad 2 \cosh(y)=a m^2-2.\]
One then finds similarly that
 \[ D_i=(-1)^i{\sinh(y(i+1))\over \sinh(y)}\]
so that
\[ \Delta_\Z D=-2(1+\cosh(y))D=- am^2 D
\]
as before, which had to happen since we recover at integer values the same $D_n$ as in (\ref{Dn}). 
We similarly have (\ref{green}) again since $D_iD_{n-2-i}-D_{i+1}D_{n-1-i}=-D_n$ holds.

This time all correlators are finite as $n\to \infty$. Indeed, for large $i$ we have 
\[ D_i\sim (-1)^i{e^{y(i+1)}\over 2 \sinh(y)},\quad \<\phi_i\phi_j\>\sim  -\imath (-1)^{i-j}{e^{y|(i-j)|}\over 2\sinh(y)}.\]
For example,  at the midpoint when $n$ is odd and $n\to\infty$,
\[ \<\phi_{n-1\over 2}^2\>=-\imath\frac{\tanh (y ({n+1\over 2}))}{ 2 \sinh(y)}\to -{\imath\over  2\sinh(y)}. \]
Note that the physics of this phase of the theory is less clear since the eigenvalues of $B_n$ from (\ref{Bn}) are bounded by $4$ so the classical theory restricted to $\phi_0,\cdots,\phi_{n-1}$ has no solutions to the wave equation for $a m^2>4$ (there are no on-shell particles). However, the functional integrals still make sense as above. 

\subsection{Plane waves}\label{secplane} We continue with the constant metric in the continuum phase where $a m^2< 4$. Here classical real plane wave solutions of the Klein-Gordon equation ${\Delta\over 2}\phi = m^2\phi$ or $\Delta_\Z\phi=-a m^2\phi$  are of the form
\begin{equation}\label{wavesol} \phi(j)= \alpha e^{-{\imath} m_0 j\sqrt{a}}+ \bar\alpha e^{{\imath}  m_0 j\sqrt{a}}\end{equation}
where $\Z$ is considered as the time direction sampled at times $t_j=j\sqrt{a}$ (as a $1+0$-dimensional scalar field theory) and 
\[  \sin\left({m_0\sqrt{a}\over 2}\right)={m \sqrt{a}\over 2},\quad 0< m_0<{\pi\over\sqrt{a}}.\]
Comparing with our previous $\sin^2({x\over 2})=(1-\cos(x))/2=a m^2/4$, we see that $m_0=x\sqrt{a}$ in terms of our previous parameter $x$. 

Here $m,m_0$ are continuous; if we restrict to modes with support $i=0,\cdots,n-1$ then we have noted (\ref{Bn}) that the eigenvalues are similar but with $m_0$ quantised. Namely the real plane waves that vanish at $i=-1,n$ have a basis
\[ \phi^{(k)}(j)=\sin({m^{(k)}_0 \sqrt{a}  (j+1)});\quad k=1,\cdots,n;\quad m_0^{(k)}={\pi k\over \sqrt{a}(n+1)},\quad m^{(k)}={2 \over \sqrt{a}}\sin({\pi k\over 2(n+1)}).\]
These can be viewed as particular solutions on $\Z$ as in (\ref{wavesol}) with $\alpha=\imath e^{-\imath m_0\sqrt{a}}/2$ or viewed as extended by zero outside of $i=0,\cdots,n-1$ in view of the vanishing boundary conditions at $i=-1,n$, which is our previous point of view. The
function $D(i)=D_i$ is exactly one of these basis functions whenever $m=m^{(k)}$ for some $k=1,\cdots,n$,  which is equivalent to $x=x^{(k)}=\pi k/(n+1)$ for some $k$. Note that in the preceding section we exactly avoided these mass values as they are the ones where $D_n$ vanishes. The correlation functions diverge for these special values of $m$ for this reason, for example the $n=2$ correlators displayed above have poles at $a\sqrt{m}=1,\sqrt{3}$.

 Back with generic $a m^2$, we close with a comparison with the conventional Hamiltonian quantisation of $\phi(j)$ in the free field case. Here
\begin{equation}\label{Phi} \Phi(j)=A e^{-{\imath} m_0 j\sqrt{a}}+ A^\dagger e^{{\imath} m_0 j\sqrt{a}}=e^{\imath H j\sqrt{a}}\Phi(0)e^{-\imath H j \sqrt{a}};\quad \Phi(0)=A+A^\dagger\end{equation}
where $[A,A^\dagger]=1$ 
and $H=m_0 (A^\dagger A+ {1\over 2})$ is the free particle Hamiltonian. The vacuum state has $A|0\>=0$ and $|n\>\propto A^\dagger{}^n|0\>$ is the $n$-particle state (normalised to unit norm). Then the correlation functions from this approach are the time ordered products 
\[ \<0| T\phi(i)\phi(j)|0\>=\<0| T\phi(j)\phi(i)|0\>= \<0|A e^{-{\imath} m_0 i\sqrt{a}} A^\dagger e^{{\imath} m_0 j\sqrt{a}}    |0\>=e^{\imath m_0(j-i)\sqrt{a}};\quad i<j \]
where $T$ denotes to put fields at earlier times to the left. Hence
\[ \<0| T\phi(i)\phi(j)|0\>= e^{\imath x |i-j|}\]
for all $i,j$. Compared to our functional integral computation, if we had $D_i\propto e^{\imath x(i+1)}$ as a complex solution of the discrete wave equation instead of its imaginary part $\imath\sin(x(i+1))$  then the correlation would have been exactly this $\<\phi_i\phi_j\>=e^{\imath x|i-j|}$. Hence our previous functional integral calculations are seeing just the imaginary part of the correlator for the full theory. This is simply a reflection of  the boundary conditions $\phi(-1)=\phi(n)=0$ imposed in the restricted functional integration whereby $\phi^{(k)}$ were a basis of eigenvectors of $B_n$ underlying the Gaussian integration, i.e. losing the cosine modes. For the full functional integration on $\Z$, one would need to  consider the full solutions (\ref{wavesol}) as a basis for diagonalisation of the underlying bilinear form, with mass parameter now a continuous variable labelling the eigenvectors (more precisely, with $x=m_0\sqrt{a}$ only relevant mod $2\pi$, i.e., a circle for the energy-momentum).

\subsection{Cosmological particle creation on curved background metrics}\label{secparticle} After our warm up with scalar field theory on a  restricted constant-metric lattice line, we now consider a general metric and the unrestricted scalar field theory over $\Z$. The general scalar wave equation is
\[  {\Delta\over 2}\phi=m^2\phi;\quad (1+\rho_{i-1})(2\phi_i-\phi_{i+1}-\phi_{i-1})= 2 a_im^2\phi_i\]
 in the explicit form, and we rewrite the latter as
 \begin{equation}\label{phievol} \phi_i=2(1-c_{i-1} m^2)\phi_{i-1}-\phi_{i-2};\quad c_i={a_{i-1} a_i\over a_{i-1}+a_i}\end{equation}
 for coefficients $c_i$ determined by the metric. Next, as in Section~\ref{secmet}, we suppose that the metric lengths $a_i=a$ are constant for $i\le 0$ and that $\rho_i=1$ for $i$ outside the range $0,\cdots,n-1$. The constant metric at later times is then $a_i=a\rho_0\cdots\rho_{n-1}=b$, say, for all $i\ge n$. In this case
\[  c_{i\le 0}= {a\over 2},\quad c_1={a a_1\over a+a_1},\quad c_2={a_{1} a_2\over a_{1}+a_2},\quad \cdots\quad c_n={a_{n-1}b\over a_{n-1}+b},\quad c_{i\ge n+1}={b\over 2}\]

Now consider for negative $i$ a real plane wave of mass $m$ of the form
\[ \phi=\alpha\bar\phi_{\rm in}+\bar \alpha\phi_{\rm in};\quad \phi_{\rm in}(i)= {e^{\imath x i}\over\sqrt{\sin(x)}},\quad i\le 1\]
as in (\ref{wavesol}) except that we have chosen a certain normalisation, which will be justified later. Here $\alpha$ is a free complex parameter and $x=m_0\sqrt{a}>0$ is such that $4\sin^2(x/2)=a m^2$ as explained in Section~\ref{secplane}. This clearly solves the wave equation for constant metric $a$ up to and including 
\[ \phi_{-1}={1\over \sqrt{\sin(x)}}\left(\alpha q+\bar\alpha q^{-1}\right),\quad \phi_0={1\over \sqrt{\sin(x)}}\left(\alpha+\bar\alpha\right);\quad q=e^{\imath x}=e^{\imath m_0\sqrt{a}}.\]
We then continue to evolve $\phi_i$ through the finite number of steps to $\phi_{n-1}$ according to (\ref{phievol}), from the form of which we see that $\phi_1$ is still governed by the initial wave equation, so
\[ \phi_1={1\over \sqrt{\sin(x)}}\left(\alpha q^{-1}+\bar\alpha q\right),\]
while $\phi_2$ starts to depart according to the value of $c_1$. Similarly, $c_{n}$ affects $\phi_{n+1}$ while  $\phi_i$, $i\ge n+2$ obey the constant plane wave equation $\phi_i=(2-b m^2)\phi_{i-1}-\phi_{i-2}$ with initial values determined by $\phi_{n+1}$ and its preceding $\phi_{n}$. Thus, similarly setting $y>0$ such that $4\sin^2(y/2)=b m^2$ for a plane wave still of mass $m$ but for metric value $b$, we match $\phi_i$ for $i\ge n$ to an outgoing wave of the form
\[ \phi=\beta \bar \phi_{\rm out}+ \bar\beta\phi_{\rm out};\quad \phi_{\rm out}(i)={e^{\imath y(i-n-1)}\over\sqrt{\sin(y)}},\quad i\ge n\]
where 
\[ {1\over \sqrt{\sin(y)}}\left(\beta+\bar\beta\right)=\phi_{n+1},\quad {1\over \sqrt{\sin(y)}}\left(\beta p+\bar\beta p^{-1}\right)=\phi_n;\quad p=e^{\imath y} \]
so that the matching is solved  by
\[ \beta=\sqrt{\sin(y)}{\phi_{n+1}- p \phi_n\over 1-p^2}=\bar f\alpha + g \bar\alpha\]
for some coefficients $ f,g\in \C$ that depend on the preceding metric values. 

Note that $\phi_{\rm in}$ separately evolves to all positive $i$  as a complex solution of  (\ref{phievol}) and $\phi_{\rm out}$ similarly evolves backwards to all smaller and negative $i$ as another complex solution. Then the above matching is equivalent to  the Bogoliubov transformation 
\[ \phi_{\rm in}=f \phi_{\rm out}+ g \bar\phi_{\rm out},\quad \phi_{\rm out}= \bar f \phi_{\rm in}- g \bar\phi_{\rm in}\]
from another point of view, with 
\begin{equation}\label{boguni} |f|^2-|g|^2=1\end{equation}
as the appropriate form of unitarity. Thus any solution $\phi$ of the wave equation can be expressed uniquely in terms of $\phi_{\rm in}$ with complex parameter $\alpha$ or uniquely in terms of $\phi_{\rm out}$ with complex parameter $\beta$, the two parameterisations being related as stated. It remains to justify the normalisations of the in and out plane waves. Normally, this is done by means of a sesquilinear inner product so that $(\phi_{\rm in},\phi_{\rm in})=1=-(\bar\phi_{\rm in},\bar\phi_{\rm in})$ and $(\phi_{\rm in},\bar\phi_{\rm in})=0$ and the same for  $\phi_{\rm out}$, where $(\ ,\ )$ is defined by a current that is conserved during evolution. This then ensures the unitarity condition (\ref{boguni}). In our case we do not yet have an understanding of conserved charges in noncommutative geometry and instead our approach is to scale the plane waves as shown so as to ensure (\ref{boguni}) directly. This fixes the  normalisation uniquely up to an overall scale as we shall see next. 

We will prove this unitarity inductively on $n$, starting with the $n=1$ case of  the metric having a single step from $a_i=a$ for $i\le 0$ to $a_i=b$ for $i\ge 1$, i.e. $n=1$ and 
\[ c_{i\le 0}={a\over 2},\quad c_1={ab\over (a+b)},\quad c_{i\ge 2}={b\over 2}.\] Then matching to an incoming plane wave with parameter $\alpha$ for negative $i$, we have
\[ \phi_2=\gamma\phi_1-\phi_0={1\over\sqrt{\sin(x)}}\left(\alpha\left(q^{-1}\gamma-1\right)+\bar\alpha \left( q\gamma-1\right)\right);\quad \gamma=2\left(1-{a b m^2\over a+b}\right). \]
and for $i\ge 3$ we have the outgoing wave equation $\phi_i=(2- b m^2)\phi_{i-1}-\phi_{i-2}$ so, following the procedure laid out above,  we match the solution to an outgoing wave with parameter $\beta$ at $i\ge 1$. This 
requires 
\[ {1\over\sqrt{\sin(y)}}\left(\beta+\bar\beta\right)=\phi_2,\quad {1\over\sqrt{\sin(y)}}\left(\beta p+\bar\beta p^{-1}\right)=\phi_1\]
and is solved by 
\begin{equation}\label{bog1} \beta={\sqrt{\sin(y)}\over (1-p^2)\sqrt{\sin(x)}}\left(\alpha(q^{-1}(\gamma-p)-1)+\bar\alpha(q (\gamma-p)-1)\right).\end{equation}
We read off $f,g$ from this and can readily check that $|f|^2-|g|^2=1$ provided the relative normalisation between the two waves is as stated. Since each plane wave has to be normalised independently, this fixes the normalisation used up to an overall constant. Note that when $b=a$, the stated expression collapses to $\beta=q^{-2}\alpha$ (noting that $\gamma=q+q^{-1}$ and $p=q$ in this case), which given the two steps difference in starting point amounts to the in and out waves coinciding in that case. 

Now consider what happens in the general case $n\ge 2$. Suppose as a virtual exercise that the metric is changed for some $0<k<n$ to have a constant value $a_{k}=b'$ for all $a_i$, $i\ge k$ and that the above matching procedure for the solution $\phi'$ results in out coefficient $\beta'=\bar{f'} \alpha+ g'  \bar\alpha$ for some $f',g'\in \C$ that depend on the preceding metric values and, which is our inductive hypothesis, have $|f'|^2-|g'|^2=1$. The out wave here has the form $e^{\imath y' (i-k-1)}/\sqrt{\sin(y')}$ for $i\ge k$ where $y'>0$ is defined by $b'm^2$. The modified metric is the same as the original $a_i$ for $i\le k$ so that the solution coincides, $\phi'_i=\phi_i$, for all $i\le k+1$. This means that the values $\phi'_k=\phi_k,\phi'_{k+1}=\phi_{k+1}$ can also be seen as matching $\phi$ to a plane wave with {\em input} coefficient $\beta'$ at $i=k+1$ (instead of $i=0$ in our previous discussion). We then continue to evolve $\phi$ with the correct metric to $i=n$ where we match as above to the true outgoing plane wave with coefficient $\beta$. Again by the inductive hypothesis, we have $\beta=\bar{f''} \beta'+  g'' \bar\beta'$ for some Bogoliubov coefficients with $|f''|^2-|g''|^2=1$. Thus the original transformation factorises and obeys 
\[f=f' f''+ g' \bar {g''},\quad g=f' g''+ g' \bar f'';\quad |f|^2 -|g|^2=(|f'|^2-|g'|^2)(|f''|^2-|g''|^2)=1.\]
One can also write the transformations here in matrix form with determinant 1. The above analysis  further tells us that the transformation from $\alpha$ at $i=0$ to $\beta$ at $i=n$ is an ordered product of $n$ 1-step transformations, i.e. some kind of holonomy of a discrete connection built from the metric (this would require a suitable formulation to elaborate further). 

Next, in view of our comments on the Hamiltonian quantisation for a free particle, we take for the quantum version a field 
\[ \Phi= A\bar\phi_{\rm in}+A^\dagger\phi_{\rm in}\]
so that $\Phi(i)$ has the plane wave form as in (\ref{Phi}) for $i\le 1$ but with a normalisation factor $1/\sqrt{\sin(x)}$. We consider that $A^\dagger$ creates $\phi_{\rm in}$ from the associated vacuum $|0\ {\rm in}\>$. However, from our later point of view we can write 
\[ \Phi=B\bar\phi_{\rm out}+B^\dagger\phi_{\rm out}\]
where $[B,B^\dagger]=1$ is another quantum harmonic oscillator. This has a similar form as our quantum free field for $i\ge n$, namely
\[ \Phi(i)= B {e^{-\imath y (i-n-1)}\over\sqrt{\sin(y)}}+B^\dagger {e^{\imath y(i-n-1)}\over\sqrt{\sin(y)}}=e^{\imath H_{\rm out}i\sqrt{b}}\Phi(n+1)e^{-\imath  H_{\rm out}i\sqrt{b}},\quad \Phi(n+1)={\rm B+B^\dagger\over\sqrt{\sin(y)}}. \]
 There is a vacuum $|0\ {\rm out}\>$ and  $H_{\rm out}={y\over\sqrt{b}}(B^\dagger B+{1\over 2})$. 
Similarly matching the two descriptions will need  $B=\bar fA +  g A^\dagger$ with the same Bogoliubov coefficients as before. This is compatible with the commutation relations for $A,B$ precisely because of the unitarity condition (\ref{boguni}). Moreover, the initial vacuum state $|0\ {\rm in}\>$ from the point of view of the later time has occupancy number
\[ \< N\> = \<0\ {\rm in}| B^\dagger B|0\ {\rm in}\>=\<0\ {\rm in}| A \bar g A^\dagger g|0\ {\rm in}\>=\<0\ {\rm in}||g|^2 AA^\dagger |0\ {\rm in}\>=|g|^2\]
by the defining properties of $|0\ {\rm in}\>$ and the $[A,A^\dagger]=1$ relation. 

For our single step $n=1$ example, we have
\begin{equation}\label{bogq} B={\imath p^{-1}\over \sqrt{-(q-q^{-1})(p-p^{-1})}}\left(A(q^{-1}(\gamma-p)-1)+A^\dagger(q (\gamma-p)-1)\right)\end{equation}
and hence the initial vacuum state from the point of view of the later time has occupancy number 
\begin{align*} \< N\>&=-{\left| q (\gamma-p)-1\right|^2\over (q-q^{-1})(p-p^{-1})}=
-{1\over (q-q^{-1})(p-p^{-1})}(2+\gamma^2-\gamma(p+p^{-1}+q+q^{-1}) + p q+ p^{-1}q^{-1})\\
&={1\over 4\sin(x)\sin(y)}(2+\gamma^2-\gamma(2\cos(x)+2\cos(y))+2\cos(x+y))\\
&={1\over \sqrt{ab(4-a m^2)(4-b m^2)}}\left(a+b-\frac{2 a^2 b m^2 (a-b)}{(a+b)^2}+ \frac{ a b m^2 (a-3 b)}{2 (a+b)}-\frac{1}{2} \sqrt{a b  \left(4-a m^2\right) \left(4-b m^2\right)}\right)\\
\end{align*}
which one can check vanishes when $a=b$. If we let $\rho=b/a$ for the fractional change in the metric at the step then
\[ \<N\>={1\over \sqrt{\rho(4- a m^2 )(4-a m^2\rho )}}\left(1+\rho-\frac{a m^2 \rho(3- 2\rho+3\rho^2)  }{2(\rho+1)^2}\right)-\frac{1}{2} \]
where $\rho$ reflects the curvature at the step and $a$ is the initial metric square length, while the mass $m$ determines the input wave frequency.  For reference, the Ricci scalar curvature for a step jump in the metric, from (\ref{RicciS}), is 
\begin{equation}\label{Sstep}S(0)=-{1\over 2a\rho}\left(\rho-1\right)=-S(2),\quad S(1)={1\over 2a\rho^2 }(\rho+1)(\rho-1)^2\end{equation}
and zero elsewhere. For $\rho>1$, this goes as $...0-+-0...$, although the reader should recall that our definition of Ricci scalar is $-1/2$ of the usual one in the continuum limit.  We see that in the continuum limit of our calculation,
\begin{equation}\label{n1limit} \lim_{a\to 0}\<N\>={1\over 4}\left(\rho^{1\over 4}-\rho^{-{1\over 4}}\right)^2\end{equation}
is necessarily independent of the frequency or mass of the input wave as $a, m^2$ always enter together as $a m^2$, which is being set to zero. Meanwhile in this limit, the jump in the metric becomes a discontinuity and the curvature diverges at the point of discontinuity. Note that we need $a m^2<4$ and $   a m^2\rho<4$ for both our input and output waves to be treated in the continuum phase of the discrete theory. The continuum limit of (\ref{phievol}) is a time-dependent harmonic oscillator of the form $\ddot\phi+c(t)m^2\phi=0$, as studied in \cite{Lew}, and our analysis indicates cosmological particle creation here as a remnant of the non-commutative curvature at least when $c(t)$ has a step discontinuity. 

The calculation for a `bump function' metric is very similar and starts off the same. Here we set metric square-length $a_i$ to be $a$ for  $i\le 0$ as before, $a_1=b$ and back to $a$ for $i\ge 2$. Hence $\rho_0=b/a=\rho$ as above and $\rho_1=a/b=1/\rho$, all others are 1. Moreover,
\[  c_{i\le 0}={a\over 2}, \quad c_1={a b m^2\over a+b}=c_2,\quad c_{i\ge 3}={a
\over 2}.\]
Then $\phi_0,\phi_1$ are just the same as before, defined by $\alpha$, and $\phi_2$ is also the same as before. The difference is 
\[ \phi_3=\gamma\phi_2-\phi_1={1\over \sqrt{\sin(x)}}\left(\alpha(q^{-1}\gamma^2-\gamma-q^{-1})+\bar\alpha(q\gamma^2-\gamma-q)\right).\]
Matching this to ${1\over\sqrt{\sin(x)}}(\beta+\bar\beta)$ for the origin of an outgoing wave, and $\phi_2$ with one step before, we have
\[ \beta={1\over 1-q^2}\left(\alpha(q^{-1}\gamma^2-2\gamma+q-q^{-1})+\bar\alpha(q\gamma^2-\gamma(1+q^2))\right).\]
In fact, the outgoing wave in the present example has the  same frequency-mass parameter $x$ as the incoming one so we could have omitted the normalisation factors. We now find
\begin{align*} \<N\>&=  \left|{q\gamma^2-\gamma(1+q^2)\over 1-q^2}\right|^2=\left|{\gamma^2-\gamma(q+q^{-1})\over q-q^{-1}}\right|^2={(\gamma(\gamma-(2-a m^2)))^2\over a m^2(4-a m^2)}\\
&={4 am^2(\rho-1)^2((1- a m^2) \rho+1)^2\over (4-a m^2)(\rho+1)^4}\sim  a  m^2\left({\rho-1\over \rho+1}\right)^2\end{align*}
 for small $a m^2$, which vanishes as $a\to 0$ even if we also send $\rho\to \infty$ in order to have a `delta-function' spike in the metric. Hence this would appear to be a purely `quantum geometry' effect (which in turn could be a quantum gravity effect presumably with $\sqrt{a}\sim \lambda_P$, the Planck scale). For reference, the Ricci scalar comes out as
\[ S(0)=S(3)=-{1\over 2 a\rho}(\rho-1),\quad S(1)=S(2)=-{1\over 2 a\rho^2}(\rho-1)(\rho^2-\rho-1).\]
which has an infinite spike in $S(1)=S(2)$ as $\rho\to \infty$ with $a$ fixed. 

The calculation for a general 2-step metric values $a_{i\le 0}=a, a_1=a \rho_0, a_{i\ge2} =a\rho_0\rho_1=b$ is very similar but with a much more complicated result, which we omit. Its continuum limit $a\to 0$, however, has occupation number exactly as in (\ref{n1limit}) for a single step but with $\rho=\rho_0\rho_1=b/a$, i.e. the ratio of the initial and final metrics and independent of the middle value.  This is also true for the general case of $n$ steps with $\rho=\rho_0\cdots\rho_{n-1}$. This is because the continuum limit $a\to0$ of the Bogoliubov transformation for a single step $a$ to $a\rho=b$ is 
\[ \beta={\alpha\over 2}(\rho^{1\over 4}+\rho^{-{1\over 4}})+{\bar\alpha\over 2}(\rho^{1\over 4}-\rho^{-{1\over 4}});\quad \beta=\begin{pmatrix}\rho^{1\over 4}&0\\ 0& \rho^{-{1\over 4}}\end{pmatrix}\alpha\]
when $\alpha,\beta$ are written as vectors in the complex plane. Assume as inductive hypothesis that the same formula applies for $m<n$ steps with $\rho_0\cdots\rho_{m-1}$ in place of $\rho$. Similarly for the remaining $n-m$ steps. Writing the relevant coefficients $f',g'$ and $f'',g''$ as diagonal matrices with entries $(\rho_0\cdots\rho_{m-1})^{\pm{1\over 4}}$ and $(\rho_m\cdots\rho_{n-1})^{\pm{1\over 4}}$, the composition $f=f' f''+ g' \bar {g''}$ and $g=f' g''+ g' \bar f''$ clearly has the same diagnonal form with $(\rho_0\cdots\rho_{n-1})^{\pm{1\over 4}}$. Armed with this result, we can now send $n\to\infty$ (so as to have a macroscopic extent of the variation region) while sending $a\to 0$ and $\rho_i\to 1$ (so as to avoid a discontinuity in the metric). Depending on how the joint limits are taken, it would appear that cosmological particle creation in the continuum will then depend only on the overall factor $\rho$ by which the metric changes  over the period of variation, and not on its shape, and still be given by (\ref{n1limit}). Indeed, it would appear that both the Ricci curvature and particle creation have continuum limit remnants as part of our new point of view on the time-dependent harmonic oscillator. 

\section{Concluding remarks}

The first and most striking conclusion is that the integer lattice admits a full moduli of quantum Riemannian geometries defined by `square lengths' attached to the edges, with most of them having curvature for the uniquely associated QLC. Key to this was to modify a previous `quantum symmetry' condition in favour of an `edge-symmetric' condition, which had also proven useful to impose in the only other known full moduli graph example\cite{Ma:sq}. This suggests that the edge-symmetric condition is more relevant and should be adopted for any graph. Having such an understanding of the geometry of the lattice line and other graphs should have many applications. We have applied it as `time' in a quantum/gravity context but one could equally well apply it for example to develop the geometry of long-chain molecules, spin-chains or other  applications completely different from theoretical physics. 

Next, for gravity, we found, remarkably, that there was a natural Einstein-Hilbert action which was based on the standard double difference wave operator, but the field that is relevant is not the metric itself but its positive-valued `relative variation' $\rho$ (where a constant metric corresponds to $\rho=1$), i.e. an exponential of the metric derivative. This relative approach was also useful in \cite{Ma:sq}. We conclude that $1+0$ dimensional quantum gravity could be viable in these variables but the integrals, albeit polynomials with Gaussian  action, are unusual due to the values of the fields and would need more study. 

For $1+0$-dimensional quantum scalar field theory in a functional integral approach on the lattice,  we conclude that, not surprisingly for a free theory, everything can be computed at least if we restrict to $n$ modes $\phi_0,\cdots,\phi_{n-1}$, but what one obtains is the imaginary part of the unrestricted theory due to the $\phi_{-1}=\phi_n=0$ boundary conditions. What was more surprising and which we have not seen elsewhere was a nice formula for the 2-point correlation function in the $n$-mode theory in terms of partition functions in the $k$-mode theory for $k\le n$.  This work was, however, intended as warm-up for the curved background theory. The latter was looked at mainly at the classical field theory level (the wave equation for the quantum-geometric $\nabla$ defined by the QLC) but assuming a Hamiltonian quantisation of the in and out waves and normalising them appropriately. We made a first calculation of cosmological particle creation effects for the 1+0 field theory with discrete curved background (or discrete time dependent harmonic oscillator). The method worked for any metric in the region of variation (after solving the discrete  wave equation through this region) and while the spectrum  was not thermal (and was unlikely to be in the absence of any space), these calculations suggest that the effect should be interesting to compute on more general curved quantum geometry backgrounds. 

It would also be interesting to compare our approach with other quantum gravity approaches, such as loop quantum cosmology (LQC)\cite{Ash, AshBar,EMM}. At the formalism level, the main difference is that in LQC, Hamiltonian quantization calculations suggest an effective modification of geometry or so-called `quantum geometry' but what the latter exactly means is unclear. By contrast we have a precise  axiomatic framework of noncommutative Riemannian geometry as suggested by the mathematics but without a clear derivation of the proposed effects from an underlying physical theory. Clearly, this would be a useful gap to bridge. In the case of LQC, one of the issues is that this uses the Ashtekar-Barbero connection and triads to describe the geometry whereas we use a more conventional metric and Levi-Civita connection approach. In fact, there is a quantum group frame bundle approach that includes moving frames and spin connections\cite{Ma:fra} but it takes rather more structure than the direct metric-connection formalism we used here. This would be necessary in any case to properly introduce fermion fields as associated to the quantum frame bundle. A different question is how our specific results compare with those of other approaches. At present, since space was only a point, we had only one in and one out harmonic oscillator and do not have an initial or final power-spectrum to compare with LQC particle creation results such as in \cite{EMM}. The situation might, however, be improved with a different graph, for example an expanding `Pascal's triangle' graph if this can be solved.  These are some directions for further work. 

We also saw in Section~\ref{secparticle} the need for an understanding of conserved currents in quantum geometry if one wants a more geometric picture behind the in and out plane wave normalisations that we used. In the continuum limit, the relevant conserved quantity is the Lewis-Riesenfeld invariant\cite{LR} of the classical time-dependent harmonic oscillator, but how this looks in terms of the quantum geometry of the integers is not clear and would be a good test for any general formalism. The Ricci curvature in our model also appeared to have some kind of remnant and role in the time dependent harmonic oscillator not visible classically. Beyond this,  it was already noted in \cite{BFV} how the Lewis-Riesenfeld invariant can motivate the construction of adiabatic vacuum states in other cosmological models such  as de Sitter spacetime. This is another direction for further work.


\begin{thebibliography}{99}
\bibitem{Dow}S. N. Ahmed, F. Dowker  and S. Surya, Scalar field Green functions on causal sets, Class. Quant. Grav. 34 (2017) 124002
\bibitem{Loll}J. Ambjorn, J. Jurkiewicz  and R. Loll, Dynamically triangulating Lorentzian quantum gravity, Nucl. Phys. B610 (2001) 347--382
\bibitem{Ash}A. Ashtekar, T. Pawlowski, P. Singh and K. Vandersloot, Loop quantum cosmology of k=1 FRW Models, Phys. Rev. D 75 (2007) 024035-1-26
\bibitem{AshBar}A. Ashtekar and A. Barrau, Loop quantum cosmology: From pre-inflationary dynamics to observations, Class. Quant. Grav. 32 (2015) 234001
\bibitem{BegMa1} E.J. Beggs and S. Majid, *-Compatible connections in noncommutative Riemannian geometry,  J. Geom. Phys. 61 (2011) 95--124
\bibitem{BegMa2}
 E.J. Beggs and S. Majid, Gravity induced by quantum spacetime, Class. Quantum. Grav. 31 (2014) 035020 (39pp)
 \bibitem{BegMa3}E.J. Beggs and S. Majid, Spectral triples from bimodule connections and Chern connections, J. Noncomm. Geom., 11 (2017) 669--701
\bibitem{BegMa4} E.J. Beggs and S. Majid, Quantum Bianchi identities and characteristic classes via DG categories, J. Geom. Phys. 124 (2018) 350--370 
\bibitem{BFV}C. Bertoni, F. Finelli and G. Venturi, Adiabatic invariants and scalar fields in a de Sitter space-time, Phys. Lett. A237 (1998) 331--336
\bibitem{Boo}A.D. Boozer, Quantum field theory in (0 + 1) dimensions, Euro. J, Phys. 28 (2007) 729--745
\bibitem{Con} A. Connes, Noncommutative Geometry, Academic Press (1994).
\bibitem{An}P.A. Damianou, On the characteristic polynomial of Cartan matrices and Chebyshev polynomials, arXiv:1110.6620 (math.RT)
\bibitem{DFR} S. Doplicher, K. Fredenhagen and J. E. Roberts, The quantum structure of spacetime at the Planck scale and quantum fields, Commun. Math. Phys. 172 (1995) 187--220
\bibitem{DVM}M. Dubois-Violette and P.W. Michor, Connections on central bimodules in noncommutative differential geometry, J. Geom. Phys. 20 (1996) 218 --232
\bibitem{EMM}B. Elizaga Navascu\'es, D. Martin de Blas and G.A. Mena Marug\'an, The vacuum state of primordial fluctuations in hybrid loop quantum cosmology, Universe 4 (2018) 98
\bibitem{FreMa}L. Freidel and S. Majid, Noncommutative harmonic analysis, sampling theory and the Duflo map in 2+1 quantum gravity, Class. Quant. Gravity 25 (2008) 045006 (37pp)
\bibitem{Lew} H. R. Lewis, Jr., Classical and quantum systems with time-dependent harmonic-oscillator-type Hamiltonians, Phys. Rev. Lett. 18  (1967) 510.
\bibitem{LR}H.R. Lewis, Jr. and W.B.  Riesenfeld, An exact quantum theory of the time-dependent harmonic oscillator and of a charged particle in a time-dependent electromagnetic field, J. Math. Phys.10 (1969) 1458--1473 
\bibitem{Ma:pla}S. Majid, Hopf algebras for physics at the Planck scale, Class. Quantum Grav. 5 (1988) 1587--1607
\bibitem{MaRue}S. Majid and H. Ruegg, Bicrossproduct structure of the $\kappa$-Poincar\'e group and non-commutative geometry, Phys. Lett. B. 334 (1994) 348--354
\bibitem{Ma:fra} S. Majid, Quantum and braided group Riemannian geometry, J. Geom. Phys. 30 (1999) 113--146
\bibitem{Ma:gra} S. Majid, Noncommutative Riemannian geometry of graphs, J. Geom. Phys. 69 (2013) 74--93
\bibitem{Ma:ltcc}S. Majid, Noncommutative differential geometry, in LTCC Lecture Notes Series: Analysis and Mathematical Physics, eds. S. Bullet, T. Fearn and F. Smith, World Sci. (2017) 139--176
\bibitem{Ma:sq}S. Majid, Quantum gravity on a square graph, arXiv:1810.10831 (gr-qc)
\bibitem{MaTao}S. Majid and W.-Q. Tao, Cosmological constant from quantum spacetime, Phys. Rev. D  91 (2015)  124028 (12pp)
\bibitem{Mou}J. Mourad, Linear connections in noncommutative geometry, Class. Quantum Grav. 12 (1995) 965 -- 974
\bibitem{MukWin}V. Mukhanov and S. Winitzki, {\em Introduction to Quantum Effects in Gravity}, Cambridge University Press (2007)
\bibitem{ParNav}L. Parker and J. Navarro-Salas, Fifty years of cosmological particle creation, arXiv:1702.07132[physics.hist-ph]
\bibitem{Sny}H.S. Snyder, Quantized space-time, Phys. Rev. D 67 (1947) 38--41
\bibitem{Wor}S.L. Woronowicz, Differential calculus on compact matrix pseudogroups (quantum groups), Commun. Math. Phys. 122 (1989) 125--170

\end{thebibliography}
\end{document}